\documentclass{emulateapj}
\usepackage{times}
\usepackage{lscape}
\usepackage{rotating}

\newcommand{\XMM}{{\em XMM-Newton }}
\newcommand{\Ch}{{\em Chandra }}
\newcommand{\Su}{{\em Suzaku }}
\newcommand{\Sw}{{\em Swift }}
\newcommand{\ot}{[O~{\sc iii}] }

\def\gappeq{\mathrel{ \rlap{\raise.5ex\hbox{$>$}}
                      {\lower.5ex\hbox{$\sim$}}  } }
\def\lappeq{\mathrel{ \rlap{\raise.5ex\hbox{$<$}}
                      {\lower.5ex\hbox{$\sim$}}  } }

\shorttitle{The Hard X-Ray View of 3C\,33}
\shortauthors{EVANS ET AL.}

\begin{document}

\title{The Hard X-Ray View of Reflection, Absorption, and the Disk-Jet Connection in the Radio-Loud AGN 3C\,33}
\author{D.~A.~Evans\altaffilmark{1,2}, J.~N.~Reeves\altaffilmark{3}, M.~J.~Hardcastle\altaffilmark{4}, R.~P.~Kraft\altaffilmark{2}, J.~C.~Lee\altaffilmark{2}, S.~N.~Virani\altaffilmark{5}}
\altaffiltext{1}{Massachusetts Institute of Technology, Kavli Institute for Astrophysics and Space Research, 77 Massachusetts Avenue, Cambridge, MA 02139}
\altaffiltext{2}{Harvard-Smithsonian Center for Astrophysics, 60 Garden Street, Cambridge, MA 02138}
\altaffiltext{3}{Astrophysics Group, School of Physical and Geographical Sciences, Keele University, Keele, ST5 5BG, UK}
\altaffiltext{4}{School of Physics, Astronomy \& Mathematics, University of Hertfordshire, College Lane, Hatfield AL10 9AB, UK}
\altaffiltext{5}{Department of Astronomy, Yale University, P.O. Box 208101, New Haven, CT 06520}

\begin{abstract}

We present results from \Su and \Sw observations of the nearby radio galaxy 3C\,33, and investigate the nature of absorption, reflection, and jet production in this source. We model the 0.5--100 keV nuclear continuum with a power law that is transmitted either through one or more layers of pc-scale neutral material, or through a modestly ionized pc-scale obscurer. The standard signatures of reflection from a neutral accretion disk are absent in 3C\,33: there is no evidence of a relativistically blurred Fe K$\alpha$ emission line, and no Compton reflection hump above 10~keV. We find the upper limit to the neutral reflection fraction is $R<0.41$ for an $e$-folding energy of 1~GeV. We observe a narrow, neutral Fe K$\alpha$ line, which is likely to originate at least 2,000~R$_{\rm s}$ from the black hole. We show that the weakness of reflection features in 3C\,33 is consistent with two interpretations: either the inner accretion flow is highly ionized, or the black-hole spin configuration is retrograde with respect to the accreting material. 

\end{abstract}


\section{INTRODUCTION}
\label{intro}

The origin of jets in active galactic nuclei (AGN) is one of the most important unsolved problems in extragalactic astrophysics. While 90\% of all AGN (Seyfert galaxies and radio-quiet quasars) show little or no jet emission, the remaining 10\% (the radio-loud AGN and radio-loud quasars) launch powerful, relativistic twin jets of particles from their cores. Since jets transport a significant fraction of the mass-energy liberated during the accretion process, sometimes out to $\sim$Mpc distances, understanding how they are produced is key to a complete picture of accretion and feedback in AGN.

X-ray observations of the nuclei of radio-loud and radio-quiet AGN are essential for establishing the connection between the accretion flow, black hole, and jet. In standard models of X-ray emission in AGN, quasi-isotropic power-law X-ray emission is generated in a corona that lies above a high $\dot{M}$ accretion disk (\citealt{kro87,ghi94}). A fraction of this hard X-ray emission is reflected by cold gas in the accretion disk, producing an observed spectrum imprinted with the features of photoelectric absorption, fluorescent 6.4 keV Fe K$\alpha$ line emission, and a Compton backscattered reflection continuum that produces a bump in the spectrum above 10 keV (\citealt{geo91}). The Fe K$\alpha$ line may be relativistically broadened, which is the signature of disk reflection near the innermost stable circular orbit (ISCO) of the black hole (e.g., \citealt{fab89}). In radio galaxies, there exists an {\it additional} power-law component of X-ray emission that is associated with the parsec-scale jet (e.g., \citealt{evans06,bal06})

X-ray observations reveal systematic differences between the spectra of radio-loud and radio-quiet AGN. Disk reflection is commonly detected in radio-quiet galaxies (\citealt{ree00,ree06}): Compton reflection bumps are prominent, and over 50\% of local Seyfert galaxies show broadened Fe K$\alpha$ emission from within 25 Schwarzschild radii. This result may hold even when the effects of complex X-ray absorption are taken into account (\citealt{nandra07}). On the other hand, radio-loud AGN tend to show weak, or absent neutral Compton reflection humps (\citealt{gra06,lar08,sam09,kat09}), and in most cases, unresolved neutral Fe K$\alpha$ lines (\citealt{bal06}).

Continuum X-ray observations of radio-loud AGN have mostly been restricted to bright broad-line radio galaxies (BLRGs) and quasars, which are oriented relatively close to the line of sight with respect to the observer. In these sources, unabsorbed non-thermal emission from the jet could potentially contaminate the unabsorbed accretion-related X-ray spectrum and thus dilute the apparent strength of the Compton reflection component. {\it Narrow}-line radio galaxies (NLRGs), on the other hand, which are oriented at low to intermediate angles, have the distinct advantage that (unabsorbed) jet-related X-ray emission can be readily spectrally separated from (heavily absorbed) accretion-related emission, allowing a direct measurement of the strength of Compton reflection. However, narrow-line radio galaxies tend to be relatively faint X-ray sources compared to broad-line objects, meaning that \Su observations of these sources are particularly useful, owing to the high effective area of the X-ray Imaging Spectrometer (XIS) and Hard X-ray Detector (HXD) instruments.

Here, we present the results from \Su and \Sw observations of one of the brightest NLRGs, 3C\,33,. It is the only NLRG to show potential evidence for Compton reflection in its 2--10 keV X-ray spectrum (\citealt{kra07}). The source may also show evidence of photoionized emission lines in its X-ray spectrum (\citealt{tor09}), which would make it one of the few radio-loud AGN to do so. In this paper, we analyze the \Su and \Sw spectra (\S\ref{analysis}), and show that reflection is {\it not} required in the source once the $>$10~keV spectrum is taken into account (\S\ref{constraintsonreflection}). We perform detailed diagnostics of the Fe~K$\alpha$ bandpass (\S\ref{suz-fek}), and demonstrate that there is no evidence for an appreciably velocity broadened component to the line. We combine data from non-contemporaneous observations with {\it Suzaku}, {\it Swift}, {\it Chandra}, and {\it XMM-Newton}, in order to search for variability in the source (\S\ref{jointfitting}). We also use these data to search for emission lines in the soft X-ray spectrum (\S\ref{softsection}). We end with a detailed interpretation of the properties of absorption, reflection, and the disk-jet connection in 3C\,33 (\S\ref{interpretation}).

\section{Observations And Data Reduction}
\label{observations}

\subsection{Suzaku}

We observed 3C\,33 with \Su on 2007 December 26 (OBSID 702059010) for a nominal exposure of 100~ks. Both the X-ray Imaging Spectrometer (XIS) and Hard X-ray Detector (HXD) were operated in their normal modes. The source was positioned at the nominal aimpoint of the HXD instrument. The data were processed using v. 2.1.16 of the \Su processing pipeline, which includes the latest Charge Transfer Inefficiency (CTI) correction applied for the XIS. We used the standard cleaned events files, which are screened to remove periods during which the satellite passed through the South Atlantic Anomaly (SAA), had a pointing direction $<$5$^\circ$ above the Earth, or had Earth day-time elevation angles $<$20$^\circ$. We describe in detail our analysis of the XIS and HXD data below.

\subsubsection{XIS}

We used data from the two operational front-illuminated (FI) CCDs (XIS0 and XIS3), together with the back-illuminated XIS1 detector. The three XIS CCDs were operated with a frame time of 8~s. For our analysis, we used data taken in the 3$\times$3 and 5$\times$5 edit modes. We selected only events corresponding to grades 0, 2, 3, 4, and 6, and removed hot and flickering pixels with the {\sc cleansis} tool.

We extracted the spectrum of 3C\,33 from the XIS CCDs using a source-centered circle of radius 2.5$'$, with background sampled from an adjacent region free from any unrelated sources, as well as the $^{55}$Fe calibration sources at the corners of each detector. We generated response matrix files (RMFs) for each detector using v. 2007-05-14 of the {\sc xisrmfgen} software, and ancillary response files (ARFs) using v. 2008-04-05 of the {\sc xissimarfgen} software. 

The net exposure times and count rates for the three CCDs are shown in Table~\ref{obslog}. We co-added the two FI spectra using the {\sc addascaspec} program, and grouped the resulting spectrum and that of the XIS1 detector to a minimum of 200 counts per bin in order to use $\chi^2$ statistics. We restricted the energy range for our spectral fitting to 0.5--10 keV.

\subsubsection{HXD}

\begin{figure}
\includegraphics[height=8cm,angle=270]{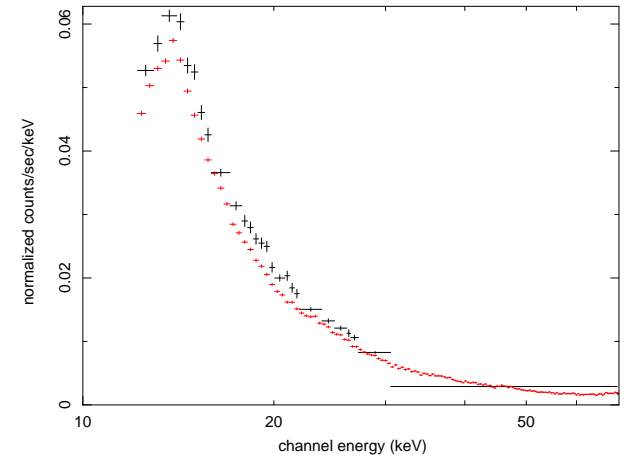}
\caption{\Su HXD/PIN spectrum of 3C\,33 {\it (black)} and total CXB+NXB background spectrum ({\it red}). The source spectrum has been binned to 3$\sigma$ above the background level.}
\label{pin_binned+background}
\end{figure}

\begin{figure}
\includegraphics[height=8cm,angle=270]{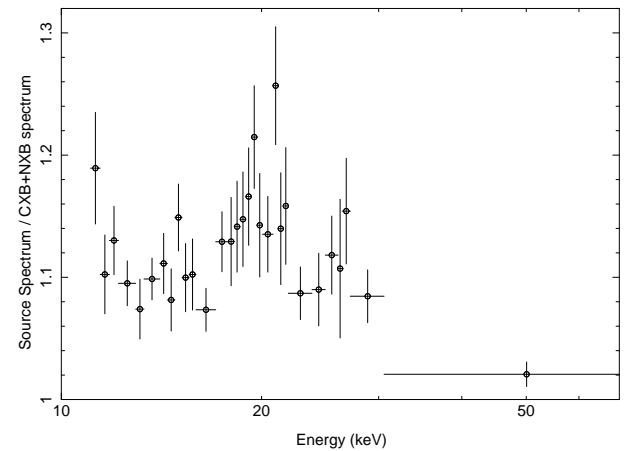}
\caption{Ratio of \Su HXD/PIN spectrum of 3C\,33 to the total CXB+NXB background spectrum. Both spectra have been binned using the 3$\sigma$ criterion described in the text.}
\label{pinratio}
\end{figure}

We extracted the source spectrum from the HXD/PIN detector, using the cleaned PIN events files described above. The source was not detected with the GSO instrument. The PIN non X-ray background (NXB) spectrum was generated from the latest `tuned' time-dependent instrumental background event file provided by the \Su Guest Observer Facility. We extracted the source and background spectra using a common Good Time Interval (GTI) criterion. The source spectrum was also dead-time corrected using the {\sc hxddtcor} tool. 

We estimated the contribution from the Cosmic X-ray Background (CXB) using the epoch-dependent PIN response for a flat emission distribution. We used a typical CXB spectrum found using HEAO-1 observations (\citealt{boldt87}) of the form $9.412 \times 10^{-3} \times (E/1~\mathrm{keV})^{-1.29} \times \exp{(-E/40~\mathrm{keV})}$~photons~cm$^{-2}$~s$^{-1}$~FOV$^{-1}$~keV$^{-1}$, where $E$ is the photon energy. We added the NXB and CXB spectra together using the {\sc mathpha} tool. Using the appropriate response file for epoch 4 of \Su observations, {\rm ae\_hxd\_pinhxnome4\_20080129.rsp}, we binned the source spectrum to 3$\sigma$ above the total (NXB + CXB) background level. The binned source spectrum and total background spectrum are shown in Figure~\ref{pin_binned+background}. In Figure~\ref{pinratio}, we plot the ratio of the source spectrum to the NXB + CXB spectrum, using our 3$\sigma$ binning criterion. It can be seen that the last bin is only marginally significant. We therefore restrict our subsequent spectral analysis to 15--30 keV.

\subsection{Swift BAT}

3C\,33 is detected in the {\it Swift} BAT 22-month all-sky survey (\citealt{tue99}). We used the preprocessed, background-subtracted BAT spectrum and standard diagonal response matrix made available to the community\footnote{http://swift.gsfc.nasa.gov/docs/swift/results/bs22mon/}. The uncertainties are calculated from the RMS noise, and include an additional multiplicative factor to account for systematic noise, as described by \cite{tue99}. Only the first 6 of the 8 total channels are above the background, so we discarded the final two bins, leaving an energy range for our spectral fitting of 14--100 keV.

\subsection{Chandra}

3C\,33 was observed for 20 ks (PI S.~S.~Murray) with the \Ch ACIS CCD camera on 2005 November 08 (OBSID 6910), and for another 20 ks on 2005 November 12 (OBSID 7200).  Both observations were performed in FAINT mode, with the source placed near the standard aimpoint of the S3 chip.  We reprocessed the data using {\sc CIAO} v4.1.2 with the CALDB v4.1.4 to create a new level-2 events file filtered for the grades 0, 2, 3, 4, and 6 and with the 0.5$''$ pixel randomization removed. To check for periods of high background, we extracted lightcurves for the entire S3 chip, excluding point sources. No periods of high background were found in either data set. The observation log is shown in Table~\ref{obslog}.

We extracted nuclear spectra from both \Ch data sets using the {\sc CIAO} routine {\sc psextract}.  We chose a source-centered region of radius $5''$, and sampled background from a surrounding annulus of inner radius $5''$ and outer radius $10''$. Finally, we grouped the spectra to 25 counts per bin. 

\subsection{XMM-Newton}

\XMM observed 3C\,33 on 2004 January 21 with the EPIC CCDs for 9 ks (PI M.~J.~Hardcastle). We reprocessed the \XMM data using {\sc SAS} version 5.4.1, and generated calibrated event files using the {\sc EMCHAIN} and {\sc EPCHAIN} scripts. We imposed the additional filtering criteria of selecting events with only the PATTERN$\leq$4 and FLAG=0 attributes. We searched for periods of high particle background by extracting light curves from the whole field of view, excluding a circle centered on the source, and selected only events with PATTERN=0 and FLAG=0 attributes and over an energy range of 10--12 keV (MOS) cameras and 12--14 keV (pn). No intervals of high background were found.

For our spectral analysis we used data from the EPIC pn camera only, owing to its greater effective area with respect to the MOS cameras. We extracted the source spectrum from circle of radius 35$''$ and extracted the background spectrum from a large off-source region on the same chip. We grouped the resulting spectrum to 30 counts per bin. Finally, we generated corresponding RMF and ARF files using the {\sc rmfgen} and {\sc arfgen} tools.

\section{Spectral Analysis}
\label{analysis}

\subsection{Overview and Motivation of Models}

In this section, we describe our spectral fitting to the broad-band X-ray spectrum of 3C\,33. Since the main goal of this paper is to constrain the properties of Compton reflection in 3C\,33, we only consider the \Su and \Sw data sets here. We introduce the \Ch and \XMM spectra in \S\ref{jointfitting}, when we perform time-variability analysis of the continuum and Fe~K$\alpha$ line.

We briefly highlight our spectral fitting methods here. In Section~\ref{singlezones}, we show that models with single zones of neutral absorption ({\bf Models I and II}) fail to provide a good fit to the spectrum of 3C\,33. This motivates us to introduce more complex spectral models, such as multiple zones of neutral absorption ({\bf Models III and IV}, described in \S\ref{multiplezones}) and a partially ionized absorbing medium ({\bf Models V and VI}, described in \S\ref{warmabs}). Once a baseline model has been established, we attempt to constrain the properties of Compton reflection ({\bf Models VII and VIII}, \S\ref{constraintsonreflection}). The casual reader may simply wish to consult Table~\ref{spectralresults} for a summary of the best-fitting models.

In all spectral fits, we linked the parameters of the model components across the \Su XIS FI, XIS BI, HXD PIN, and \Sw BAT spectra. We tied the normalizations of the FI and BI XIS spectra together, but applied a constant cross-normalization factor of 1.09 for the PIN spectrum relative to the XIS as described in the \Su Data Reduction Guide\footnote{http://heasarc.gsfc.nasa.gov/docs/suzaku/analysis/abc/}. We also included a Gaussian of initial centroid 6.4 keV to represent the Fe~K$\alpha$ emission, and allowed its energy, line width and normalization to vary freely. We report in detail the properties of the Fe K$\alpha$ line in \S\ref{suz-fek}.

We performed our spectral fitting using v.1.5.0-13 of the {\sc ISIS} spectral fitting package\footnote{http://space.mit.edu/cxc/isis/}, which has been modified to run on a highly parallelized compute platform. All results presented here use a cosmology in which $\Omega_{ m, 0}$ = 0.3, $\Omega_{\rm \Lambda, 0}$ = 0.7, and H$_0$ = 70 km s$^{-1}$ Mpc$^{-1}$. At the redshift of 3C\,33, $z$=0.0597, the luminosity distance is 267.3~Mpc. Errors quoted are 90 per cent confidence for one parameter of interest (i.e., $\chi^2_{\rm min}$ + 2.7), unless otherwise stated. All spectral fits include the Galactic absorption to 3C\,33 of $N_{\rm H, Gal}=3.03\times10^{20}$ cm$^{-2}$ (\citealt{kal05}).

\begin{figure}
\includegraphics[height=8cm,angle=270]{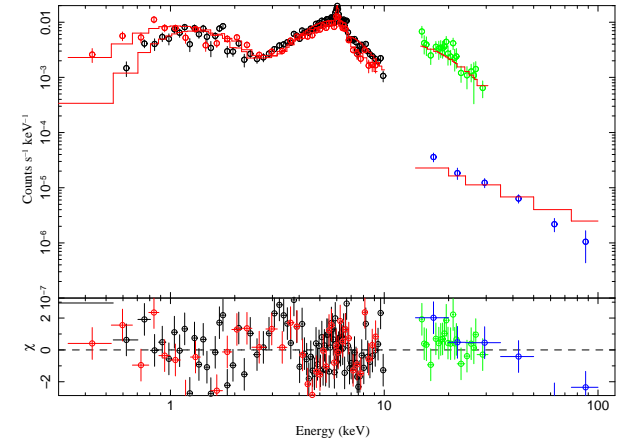}
\caption{\Su XIS FI {\it (black)}, XIS BI {\it (red)}, HXD PIN {\it (green)}, and \Sw BAT {\it (blue)} spectra and residuals of 3C\,33. The model fit is the sum of a heavily absorbed power law, a soft, unabsorbed power law (to take into account the soft excess), and a Gaussian Fe~K$\alpha$ emission line {\bf (Model I)}. The noticeable residuals at $\sim$3~keV motivates us to consider more complicated spectral models.}
\label{model1}
\end{figure}

\subsection{Single Zones of Neutral Absorption: Models I and II}
\label{singlezones}

We first fitted to 3C\,33 the canonical model for the X-ray spectrum of NLRGs: the combination of a heavily absorbed power law and a soft, unabsorbed power law  (\cite{evans06,bal06}) {\bf (Model I)}. We tied together the photon indices of both power laws. The model resulted in a poor fit to the spectrum ($\chi^2=242$ for 152 dof), with noticeable residuals at energies $\sim$3~keV (see Fig.~\ref{model1}), and an atypically low photon index ($\Gamma$=$1.39\pm0.05$). An identical fit is, of course, achieved if we allow the power law to be partially covered by neutral material {\bf (Model II)}; in this case, the covering fraction is $96\pm1$\%.

\subsection{Multiple Zones of Neutral Absorption: Models III and IV}
\label{multiplezones}

\begin{figure}
\includegraphics[height=8cm,angle=270]{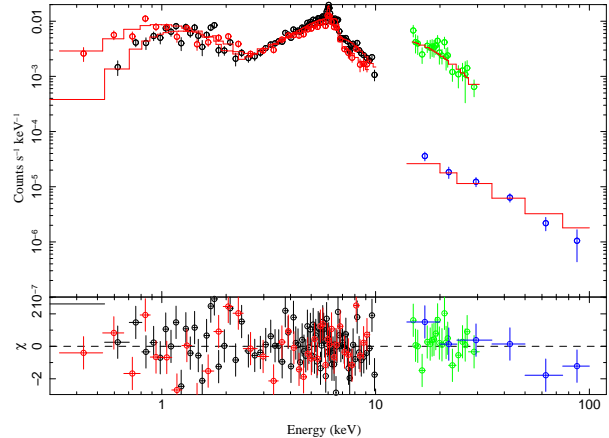}
\caption{As Fig.~\ref{model1}, but fitted to a double partially covered primary law {\bf (Model III)}. This model provides a good fit to the spectrum ($\chi^2=176$ for 150 dof).}
\label{model3}
\end{figure}

We next investigated the possibility that an additional layer of cold absorption is required to fit the broadband spectrum of 3C\,33 adequately. We adopted a double partial-covering model, implemented in {\sc XSPEC/ISIS} as \textrm{phabs$\times$zpcfabs(1)$\times$zpcfabs(2) $\times$(powerlaw+zgauss)}, where {\sc zpcfabs} represents the partial coverer ({\bf Model III}). The individual column densities $N_{\rm H, 1}$ and $N_{\rm H, 2}$, respectively have covering fractions $f_1$, $f_2$. 

Table~\ref{spectralresults} shows the best-fitting parameters for this model, and Figure~\ref{model3} shows the counts spectrum, model, and residuals. In short, the model provides a good fit to the spectrum ($\chi^2$=176 for 150 dof), and represents a substantial improvement over Models I and II. Further, the photon index of the power law ($\Gamma_{\rm 1}=1.71\pm0.09$) is significantly steeper than in the previous model, and is now consistent with typical values found in narrow-line radio galaxies (e.g., \citealt{bal06}). Again, an identical fit is found if we choose not to explicitly associate the heavily absorbed and unabsorbed spectral components  {\bf (Model IV)}.

\subsection{Warm Absorption: Models V and VI}
\label{warmabs}

An alternative method of modeling the observed spectral complexity in 3C\,33 is to allow the absorber to be partially ionized: such an absorber transmits more of the low-energy continuum emission than does a neutral obscurer of identical column density. We modeled the \Su spectrum as a power law, partially covered by a single zone of partially photoionized absorption ({\bf Model V}). The ionized absorbed is modeled by the {\sc Zxipcf} model. 

\begin{figure}
\includegraphics[height=8cm,angle=270]{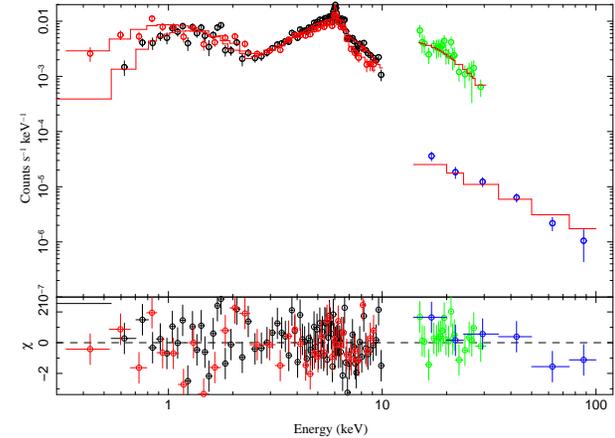}
\caption{As Fig.~\ref{model1}, but fitted to a power law partially covered by a single zone of warm absorption {\bf (Model V)}. The model provides a good fit to the spectrum ($\chi^2=184$ for 151 dof).}
\label{model5}
\end{figure}

This model provided a good fit to the spectrum ($\chi^2=184$ for 151 dof), with a column density $N_{\rm H}=(5.2^{+0.3}_{-0.2})\times10^{23}$~cm$^{-2}$ of modestly ionized gas [log$(\xi)$=$1.51^{+0.17}_{-0.20}$ ergs~cm~s$^{-1}$] as its best-fitting absorption parameters. The spectrum and model are shown in Figure~\ref{model5}. Again, an identical fit is found if we choose not to explicitly associate the heavily absorbed and unabsorbed spectral components  {\bf (Model VI)}.

\subsection{Definition of baseline model}

We have demonstrated that the broadband X-ray spectrum of 3C\,33 can be described with a power law that is partially covered by multiple zones of neutral absorption {\bf (Model III)}, or by a power law that is modified by a partially-covering warm absorber {\bf (Model V)}. Both fits include neutral Fe~K$\alpha$ emission. We have also shown that {\it mathematically identical fits} are achieved if we do not associate the hard and soft X-ray components with the same physical process {\bf (Model IV and Model VI)}. At this stage, we define as our baseline best-fitting spectral models a Model IV, since it is the most statistically acceptable. We emphasize that this does not change the results of our spectral fitting in subsequent sections. We provide a detailed discussion of the merits of both the neutral and ionized absorption models in \S\ref{interpretation}.

\section{Constraints on Reflection}
\label{constraintsonreflection}

We determined the strength of any Compton reflection in the spectrum of 3C\,33 by replacing the primary power law in our baseline model with a {\sc pexrav} component ({\bf Model VII}). This model calculates the sum of the direct power-law emission plus the fraction, $R$, reflected from an infinite slab of neutral material. We initially froze the $e$-folding energy of the {\sc pexrav} component at 1000~keV (i.e., we chose there to be no spectral cutoff in our \Su spectrum), adopted an inclination angle of 60$^{\circ}$, and fixed the elemental abundances at their solar values.

\begin{figure}
\includegraphics[height=8cm,angle=270]{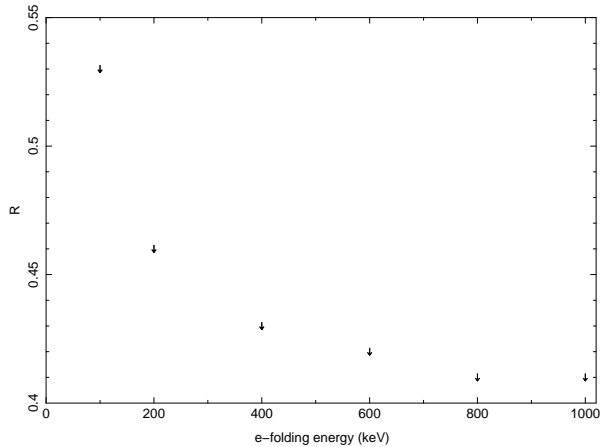}
\caption{90\% confidence upper limit to the neflection fraction as a function of e-folding energy.}
\label{r_vs_efolding}
\end{figure}

This neutral reflection model provided no significant improvement in the fit ($\chi^2$=175 for 149 dof). The best-fitting value of the reflection fraction is $R=0$, with a 90\% confidence upper limit of $R<0.41$. The column densities, covering fractions, and power-law photon index are all consistent at the 1$\sigma$ level with the values in Model IV. The best-fitting parameters are given in Table~\ref{spectralresults}. Since the strength of reflection correlates with the e-folding energy, we repeated our fits with different values of $E_{\rm cut}$, ranging from 100 to 1000 keV. We show the upper limits to $R$ as a function of e-folding energy in Figure~\ref{r_vs_efolding}.

We also attempted to fit the spectrum with an {\it ionized} Compton reflection model, using the the {\sc Reflionx} table model (\citealt{rf05}) ({\bf Model VIII}). This model resulted in a similar fit to the spectrum ($\chi^2$=174 for 148 dof), and the ionization parameter pegged at its hard limit of $\xi$=$10^{4}$ ergs~cm~s$^{-1}$.

\section{Fluorescent Line Emission}
\label{suz-fek}

\begin{figure}
\includegraphics[height=8cm,angle=270]{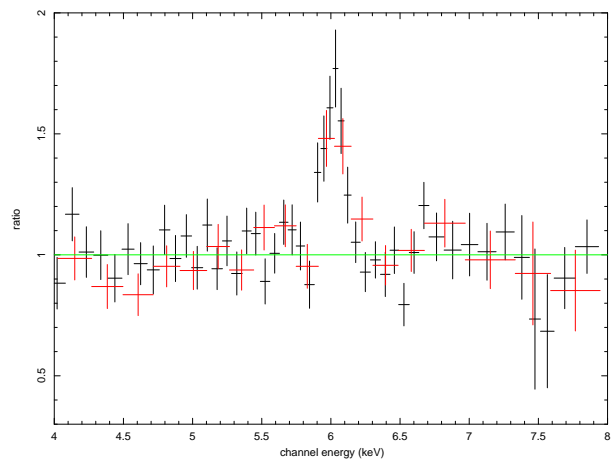}
\caption{Data/model residuals for \Su XIS FI {\it (black)} and XIS BI {\it (red)} spectra in the energy range 4--8 keV, after fitting Model IV to the FI, BI, and PIN spectra to the broadband spectrum, but excluding the energy range 6.2--6.5 keV (rest frame). Prominent residuals at $\sim$6.4~keV are observed, indicating the need for a narrow Fe~K$\alpha$ line. No red wing to the Fe~K$\alpha$ emission is observed.}
\label{nofek}
\end{figure}

\begin{figure}
\includegraphics[height=8cm,angle=270]{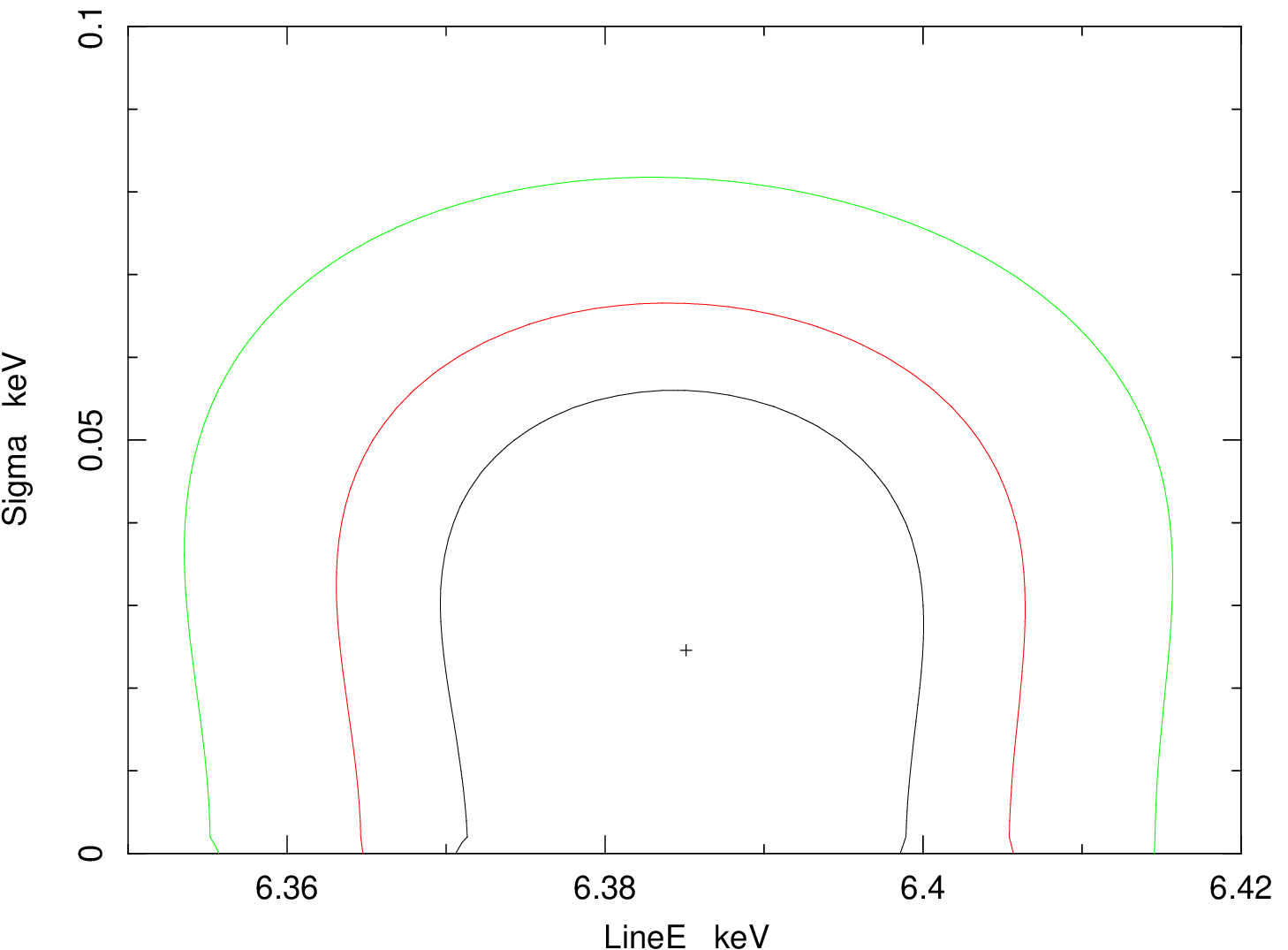}
\caption{Energy and line width confidence contours for the Fe~K$\alpha$ line measured from the \Su XIS FI and BI spectra. Shown are the 68\%, 90\%, and 99\% contours.}
\label{fek_conf}
\end{figure}

We next measured the properties of the fluorescent Fe~K$\alpha$ emission. To give a {\it qualitative sense} of the line profile, we fitted the continuum-only components of Model IV to the XIS FI, XIS BI, and HXD PIN spectra in the energy range 2--70 keV. We excluded the rest frame Fe~K$\alpha$ bandpass of 6.2--6.5 keV. The continuum parameters of this fit are consistent at the 1$\sigma$ level to those found in Model IV. We then extended the energy range to include the Fe~K$\alpha$ bandpass. The data/model residuals between 4 and 8 keV are shown in Figure~\ref{nofek}. There are clear residuals at a rest-frame energy of 6.4~keV, corresponding to neutral Fe~K$\alpha$ emission.

Our best-fitting model (Model IV), which we discussed in \S\,\ref{multiplezones} includes Fe~K$\alpha$ emission with the following properties: its centroid is $6.385^{+0.021}_{-0.023}$~keV and the upper limit to the line width is $<$65~eV (both values are quoted at 90\% confidence for two interesting parameters). This is less than the $\sim$100~eV FWHM resolution of the XIS at $\sim$6~keV. Confidence contours for the energy and line width are shown in Figure~\ref{fek_conf}. The equivalent width of the Fe~K$\alpha$ line is $129\pm43$~eV. The parameters of the Fe~K$\alpha$ line are summarized in Table~\ref{fek_parameters}, and are consistent with those found with \Ch and \XMM observations (\citealt{kra07,evans06}). The addition of a Fe~K$\beta$ line is statistically insignificant.

There is no indication of a velocity broadened component of the Fe~K$\alpha$ emission, either in the form of a broad underlying Gaussian or a relativistically blurred `diskline'. However, given that some radio galaxies may have modestly broadened, {\it ionized} Fe~K$\alpha$ lines (e.g.,~\citealt{kat07}), we calculated the upper limit to the equivalent width of a Gaussian of frozen line width $\sigma$=0.5~keV and energy 6.8~keV. We found the 90\% confidence upper limit to be $<$65~eV.

\begin{figure}[b]
\includegraphics[height=8cm,angle=270]{f9.eps}
\caption{Counts spectra, best fitting model, and residuals for all data. Shown are the \Su XIS FI {\it (black)}, XIS BI {\it (red)}, HXD PIN {\it (green)}, \Sw BAT {\it (blue)}, \Ch OBSID 6910 {\it (light blue)}, \Ch OBSID 7200 {\it (magenta)}, and \XMM pn {\it (orange)}. The model fit shown is the sum of a partially covered power law, a soft, unabsorbed power law, and a Gaussian Fe~K$\alpha$ line (Model IV).}
\label{jointplot}
\end{figure}

\begin{figure*}[t]
\centering
\includegraphics[width=16cm,angle=0]{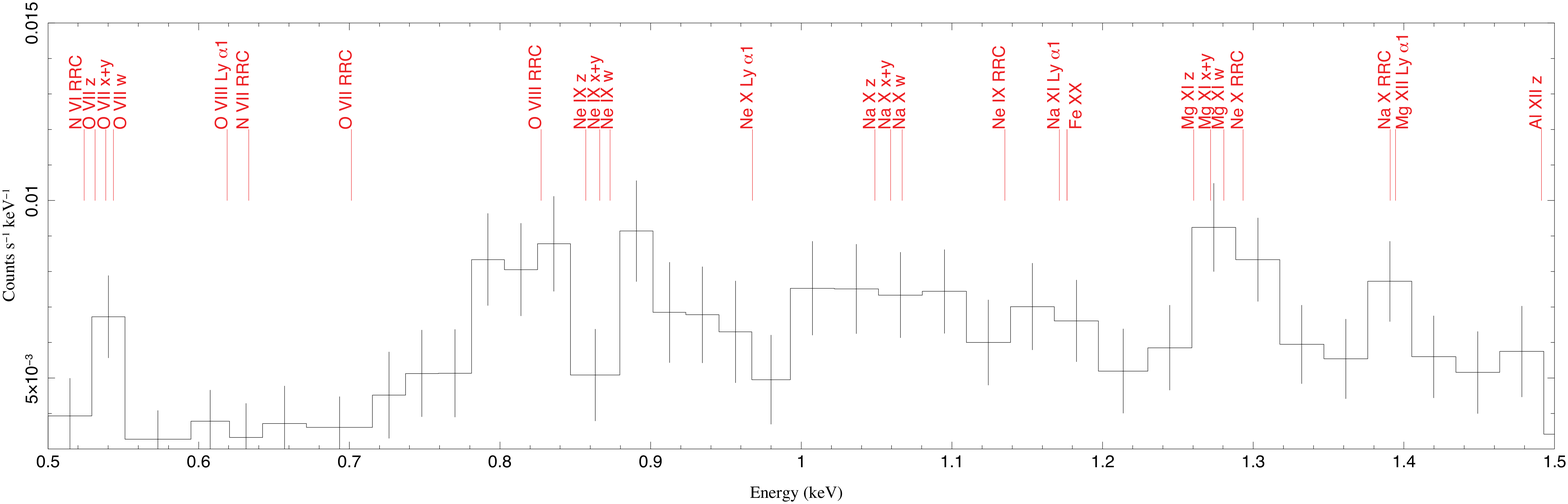}
\caption{Combined {\it Chandra}, {\it XMM-Newton}, and {\it Suzaku} CCD spectrum of 3C\,33 in the energy range 0.5--1.5 keV. Also plotted are the spectral transitions identified in the photoionized plasma of the canonical Seyfert 2 galaxy, NGC~1068.}
\label{softlines}
\end{figure*}

\section{Searching for Variability: Joint Fitting to Suzaku, Swift, Chandra, and XMM-Newton Spectra}
\label{jointfitting}

We searched for inter-epoch spectral variability of 3C\,33 by simultaneously fitting the {\it Suzaku} XIS FI, PI, and PIN, {\it Swift} BAT, {\it Chandra} ACIS (two observations), and {\it XMM-Newton} EPIC/pn spectra. We adopted our baseline fit (Model IV) for our spectral fits. We used this model instead Model III because the normalizations of the soft power laws are likely to differ considerably. This is due to the different extraction regions used for our {\it Suzaku} XIS (2.5$'$), {\it Chandra} (5$''$), and {\it XMM-Newton} (35$''$) analyses, which sample dissimilar amounts of extended, soft X-ray emission. We restricted the energy range of the fits to 0.5--10 keV (\Su XIS FI and BI), 15--30~keV (\Su PIN), 14-100 keV ({\it Swift} BAT), 0.5--7 keV ({\it Chandra}), and 0.5--8 keV ({XMM-Newton}).

For our spectral fitting, we linked the column densities, covering fraction and photon index associated with the primary power law across all five data sets, and tied the photon index of the unabsorbed power law to that of the absorbed component. We allowed the normalizations of the {\it Swift}, {\it Chandra}, {\it XMM-Newton}, and (combined) \Su power laws to vary independently, and applied the \Su PIN/XIS cross-normalization factor of 1.09. Finally, we added an unresolved Gaussian Fe~K$\alpha$ line to our model.

The model provided a good description of all the spectra ($\chi^2 = 292$ for 261 dof). All the parameters are shown in Table~\ref{jointresults}. In summary, the best-fitting parameters of the partially covered power law are $N_{\rm H}$=$(4.9^{+0.6}_{-0.5})$$\times10^{23}$~cm$^{-2}$ with a covering fraction of $92^{+3}_{-4}$~\%, and photon index $\Gamma$=$1.67\pm0.10$. The unabsorbed 2-100~keV (extrapolated) luminosities of the primary power law are consistent with no inter-epoch variability. In Figure~\ref{jointplot} we show the counts spectra and model residuals for the seven datasets.

\section{Soft-Band X-Ray Emission}
\label{softsection}

In order to search for emission lines that might indicate emission from a thermal plasma, as first reported by \cite{tor09}, we combined the {\it Chandra}, {\it XMM-Newton}, and {\it Suzaku} spectra together using the {\sc ISIS} spectral fitting package\footnote{http://space.mit.edu/cxc/isis/}. We rebinned each data set to a common grid, namely the HWHM of the \Su XIS CCDs. In Figure~\ref{softlines}, we show the fluxed spectrum between 0.5 and 1.5 keV, and plot the key photoionization transitions identified in the \Ch HETGS spectrum of the canonical Seyfert 2 galaxies NGC~1068 (\citealt{ogle03}) and NGC~2110 (\citealt{evans06}). We interpret these results in \S\,\ref{interp-soft}.

\section{Interpretation}
\label{interpretation}

We have shown that the nuclear X-ray spectrum of 3C\,33 is well described by a power law that is transmitted either through one or more layers of neutral material, or through a partially ionized obscurer. There is no evidence for a Comptonized reflection continuum above $\sim$10~keV, which means we have now resolved the question of reflection raised by previous \Ch observations (\citealt{kra07}). In this section, we use the Fe K$\alpha$ line to diagnose the physical location and state of the fluorescent emission (\S\ref{interp-fek}), discuss the origin of the intrinsic absorption in 3C\,33 (\S\ref{interp-absorption}), consider various origins for the soft X-ray emission (\S\ref{interp-soft}), and finally interpret the accretion, reflection and possible jet-production properties of 3C\,33 in the context of other radio-loud and radio-quiet AGN (\S\ref{interp-reflection}).

\subsection{Fe K$\alpha$ Diagnostics}
\label{interp-fek}

The strong Fe~K$\alpha$ line detected with \Su allows us to place constraints on the location and physical state of the fluorescing material. The energy of the line core, $6.385^{+0.021}_{-0.023}$~keV, is consistent with Fe fluorescence from neutral or near-neutral (up to $\sim$Fe~{\sc XVIII}) species. The unresolved Fe K$\alpha$ line width and the estimated black hole mass of $\sim$$4\times10^{8}$ M$_\odot$ (\citealt{smith90,bettoni03}) provide a lower limit to the emission radius of $\gappeq0.1$~pc ($\sim$2000~R$_{\rm s}$), using Keplerian arguments. Finally, the equivalent width of the Fe~K$\alpha$ line is consistent with transmission through an absorbing column of relative iron abundance $A_{\rm Fe}$=1.0 and column density similar to that measured from our spectral fitting (\citealt{ghi94,miy96}). 

There is no evidence for an additional velocity broadened component of the Fe~K$\alpha$ line, which rules out the presence of relativistically blurred fluorescent emission from the innermost portions of the accretion flow. This may imply that a standard, cold Shakura-Sunyaev accretion disk is truncated at a large distance from the central supermassive black hole. We cannot rule out the presence of a modestly broadened ($R\sim100R_{\rm s}$), ionized Fe~K$\alpha$ line in 3C\,33 but nonetheless it seems clear that fluorescent emission from the inner disk is absent. We will return to this point in \S\ref{interp-reflection}.

\subsection{Origin of Absorption}
\label{interp-absorption}

We considered in detail several absorption models that can account for the nuclear spectrum. In one (Model III), two zones of neutral material (a partial coverer of column $N_{\rm H, 1}$$\sim$$6\times10^{23}$~cm$^{-2}$ with a covering fraction of $\sim$80\%, plus a second column of $N_{\rm H, 2}$$\sim$$2\times10^{23}$~cm$^{-2}$) obscure the primary emission. The second column takes the form of a patchy absorber if we explicitly choose to associate the soft, unabsorbed, X-ray emission with the primary power law (Model IV). Alternatively, we can model the absorber as a single zone of modestly ionized material (Model V), which transmits more of the low-energy continuum emission than a neutral obscurer and thus reproduces the spectral complexity in the 2--10 keV spectrum. The relatively mild ionization of the warm absorber is also consistent with the neutral or near-neutral Fe~K$\alpha$ emission observed. Again, the warm absorber is patchy if we choose to associate the soft X-ray continuum with the same physical process as the primary power law.

Both models have distinct advantages over simple parameterizations of the nuclear continuum: they adequately explain the low-energy spectrum and reconcile the photon index of the primary power law with canonical values found in radio galaxies. However, we are unable to distinguish between these models due to the lack of photon statistics at lower energies, where discrete ionized absorption features would be observed in the case of the warm absorber.

If we choose to associate the unresolved fluorescent Fe~K$\alpha$ emission with the circumnuclear absorber then, as calculated above, the lower limit to the radius of the obscurer is $\sim$0.1~pc. We can also calculate the upper limit to the radius, $r$, in the case of the warm absorber model, since the ionizing luminosity, $L=\xi nr^2$, where $\xi$ is the ionization parameter and $n$ is the electron number density. Assuming that the thickness of the absorber is less than the distance to the absorber from the source of ionization, i.e., $\Delta r / r < 1$, we can establish the inequality $r < L / N_{\rm H}\xi$. We find $R\lappeq20$~pc, which is consistent with the parsec-scale absorber invoked in AGN unification models.

\subsection{Origin of Soft X-ray Emission}
\label{interp-soft}

There is clear evidence of a `soft excess' in the X-ray spectrum of 3C\,33. We consider several possibilities for its origin here. First, the emission could be the result of patchy circumnuclear absorption (either neutral or partially ionized), which allows a small fraction of continuum emission to be transmitted free from obscuration. Such a model is consistent with observations of numerous AGN, including NGC\,2110 (\citealt{evans07}). The lack of variability of the source on timescales of years (\S\ref{jointfitting}) suggests that such a patchy absorber would subtend a large solid angle at the source of the X-ray emission.

Second, the soft emission could represent the scattering of the primary (heavily obscured) power-law emission in to the line of sight, as is predicted by unified models of AGN (e.g., \citealt{up95}). Here, the emission is scattered by free electrons in the vicinity of the broad-line region and may be associated with the (possibly photoionized) circumnuclear regions of the AGN. The `efficiency' of scattering is simply given by the ratio of the normalizations of the primary and soft power laws --- for 3C\,33, we find the efficiency to be $\sim$2\%. This number is consistent with the typical range found in Seyfert galaxies of 1--10\% (\citealt{tur97}). Further support for this model comes from the spatially extended soft X-ray and \ot on scales of $\sim$5~kpc emission seen in \Ch and {\it HST} observations of the source (\citealt{kra07}). The emission is elongated along the axis of the large-scale radio emission in the source, and therefore likely along the axis of the circumnuclear obscurer described by unification models. In this case, we might expect a contribution from photoionized emission lines in the soft X-ray spectrum  in the case of 3C\,33, as argued by \cite{tor09}. Our combined soft X-ray spectrum (Fig.~\ref{softlines}) shows possible hints of He-like species of O~{\sc VII} and Mg~{\sc XI}, but there are insufficient photon statistics to determine if we are witnessing photoionized gas.

\begin{figure}
\includegraphics[width=8cm,angle=0]{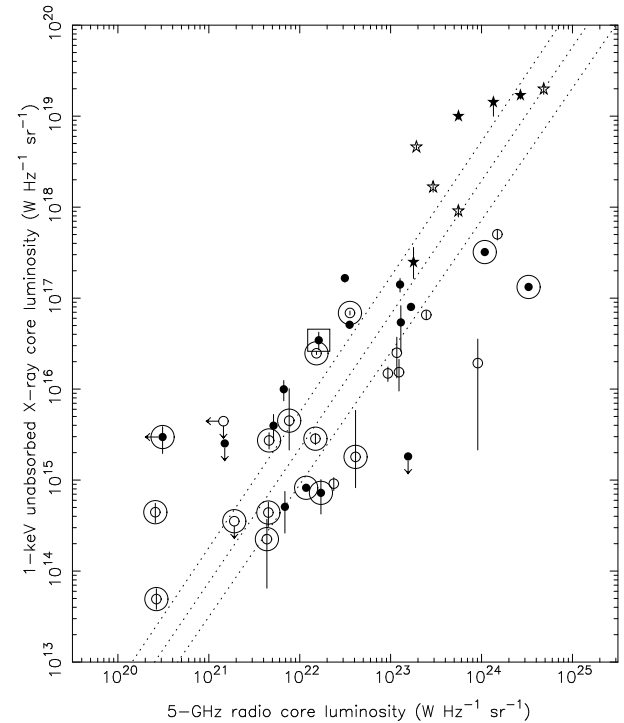}
\caption{1-keV X-ray luminosity of the soft X-ray emission component as a function of 5-GHz radio core luminosity in all observed 3CRR sources at $z<0.5$ (\citealt{evans06,hec06}). Open circles are LERG, filled circles NLRG, open stars BLRG, and filled stars quasars. Large surrounding circles indicate that a source is an FRI. 3C\,33 is marked by a box. Dotted lines show the regression line to all data and its 1$\sigma$ confidence range. The similarity of 3C\,33 to the other sources suggests that its soft X-ray component originates in an unresolved jet, although we cannot rule out other interpretations.}
\label{lum-lum}
\end{figure}

A third and final possibility is that the soft X-ray emission is {\it not} associated with the primary X-ray emission related to the accretion flow, and instead originates in an unresolved parsec-scale jet. Given that 3C\,33 is a strongly radio-loud AGN, a contribution to the X-ray emission from the jet is not unexpected, and has been argued for extensively in other radio galaxies by, e.g., \cite{har99} and \cite{bal06}. The key to these authors' interpretations is the strong observed correlation between the X-ray and radio core luminosities in large samples of radio-loud AGN, which they argue supports a common origin of the two components at the base of a jet. In Figure~\ref{lum-lum}, we show the 1-keV and 5-GHz luminosity densities of all observed $z$$<$0.5 radio galaxies (\citealt{hec06}), together with 3C\,33. This figure shows that 3C\,33 has a similar measured X-ray luminosity to that predicted from the \cite{hec06} relation.

In summary, we cannot distinguish between these three physical scenarios. Future observations with the gratings instruments on board \Ch and \XMM would allow us to determine which is correct.

\subsection{Accretion and Reflection in 3C\,33}
\label{interp-reflection}

Our results demonstrate that 3C\,33 shows no signs of Compton reflection from neutral material in the inner regions of an accretion disk: there is no reflection hump at energies $>$10~keV, and no evidence for relativistically broadened Fe~K$\alpha$ emission (Model VII). Indeed, the narrow Fe K$\alpha$ line is likely to originate in a much farther region, at least 2,000~R$_{\rm s}$ from the black hole. Further, as 3C\,33 is a {\it narrow}-line radio galaxy, meaning we can spectrally separate the X-ray emission associated with the accretion flow and jet, we rule out the idea that X-ray emission from the jet dilutes the reflection spectrum of the source at high energies. The lack of a Compton hump in 3C\,33 is consistent with growing numbers of radio-loud AGN observed with \Su to show systematically weaker reflection than their radio-quiet counterparts (e.g., \citealt{lar08,sam09,kat09}). There are two principal models that can explain this dichotomy, which basically stem from differences in either the accretion flow geometry, or the black-hole spin configuration. We examine both in the context of 3C\,33 below.

\subsubsection{Ionized Accretion Models}
\cite{bal02} interpret the weak reflection features observed in radio-loud AGN such as 3C\,33 to be the result of an ionized inner accretion flow. Since the reflected spectrum of an incident power law on to highly ionized material is itself a power law, they demonstrate that the reflection fraction, $R$, can remain arbitrarily high. Our attempt to fit an ionized reflection model (\citealt{rf05}) to 3C\,33 led to the ionization parameter pegging at its hard limit of $\xi$=$10^{4}$ ergs~cm~s$^{-1}$ (Model VIII). For 3C\,33, therefore, we cannot distinguish between a very highly ionized reflector, or simply an absent inner disk. We further note that the lack of ionized Fe~K$\alpha$ emission is consistent with this hypothesis.

\subsubsection{Retrograde Black-Hole Spin}
The X-ray spectrum of 3C\,33 is also consistent with the picture of \cite{ges09}. They argue the jet power of high-excitation radio galaxies such as 3C\,33 is the result of retrograde black-hole spin with respect to the accreting material. This configuration results in a larger innermost circular stable orbit than for a prograde black hole. The weak or absent signatures of reflection in 3C\,33, then, are simply a consequence of the large inner disk radius. Furthermore, \cite{ges09} show that the retrograde spin configuration produces both strong Blandford-Payne and Blandford-Znajek jets (\citealt{bp82} and \citealt{bz77}, respectively). They point out that this highly relativistic motion away from the inner accretion flow can easily wash out reflection features, as first demonstrated by \cite{rey97} and \cite{bel99}. Our X-ray observations of 3C\,33 support this scenario.

In summary, we have presented two, equally plausible, models that can explain the paucity of a Compton reflection hump in 3C\,33. Future, high-sensitivity observations of 3C\,33 with {\it IXO} will be needed to distinguish between them.

\section{Conclusions}
\label{conclusions}

We have presented results from a new 100-ks \Su observation of the radio galaxy 3C\,33. The \Su data have allowed us to break degeneracies in the reflection and absorption properties of 3C\,33 that were present in our previous \Ch and \XMM observations. We have shown the following:

\begin{enumerate}
\item The nuclear X-ray spectrum of 3C\,33 is well described by a power law of photon index $\sim$1.7 that is absorbed either through one or more layers of pc-scale neutral material, or through a partially ionized pc-scale obscurer. Both models have distinct advantages over simple parameterizations of the nuclear continuum: they adequately explain the low-energy spectrum and reconcile the photon index of the primary power law with canonical values found in radio galaxies.
\item There is evidence for a `soft excess' in the X-ray spectrum of 3C\,33. Its origin is consistent with several possibilities, including patchy circumnuclear absorption, photoionized gas, and an unresolved parsec-scale jet.
\item 3C\,33 shows no signs of Compton reflection from neutral material in the inner regions of an accretion disk: the upper limit to the reflection fraction for neutral material is $R<0.41$ for an $e$-folding energy of 1~GeV.
\item The observed Fe K$\alpha$ line originates in neutral or near-neutral material  at least 2,000~R$_{\rm s}$ from the black hole. We cannot rule out the presence of a modestly broadened ($R\sim100R_{\rm s}$), ionized Fe~K$\alpha$ line in 3C\,33 but nonetheless it seems clear that fluorescent emission from the inner disk is absent.
\item The absence of a Compton reflection hump in 3C\,33 is consistent with either an {\it ionized} accretion flow (\citealt{bal02}), or with retrograde black-hole spin (\citealt{ges09}). Future observations with {\it IXO} will allow us to distinguish between these two models.
\end{enumerate}

\acknowledgements

We wish to thank the anonymous referee for comments which greatly improved this paper. DAE gratefully acknowledges financial support for this work from NASA under grant number NNX08AI59G. MJH thanks the Royal Society for a Research Fellowship. This research has made use of data obtained from the Suzaku satellite, a collaborative mission between the space agencies of Japan (JAXA) and the USA (NASA). This work has also made use of the \Ch X-ray Observatory, which is operated by the Smithsonian Astrophysical Observatory for and on behalf of NASA under contract NAS8-03060. This work is additionally based on observations obtained with {\it XMM-Newton}, an ESA science mission with instruments and contributions directly funded by ESA Member States and NASA. Finally, this research has made use of the NASA/IPAC Extragalactic Database (NED) which is operated by the Jet Propulsion Laboratory, California Institute of Technology, under contract with the NASA.

\clearpage
\newpage

\begin{table}\small
\centering
\caption{{\sc Observation Log}}
\vspace{0.2cm}
\begin{tabular}{llllll}
\hline\hline
Observatory & Date of observation       & Obs. ID & Instrument         & Screened exposure time & Net nuclear count rate (s$^{-1}$)    \\ \hline
\Su         & 2007 Dec 26               & 702059010         & XIS0       & 127 ks                 & $(5.47\pm0.14)\times10^{-2}$ \\
            & 2007 Dec 26               &                   & XIS1       & 127 ks                 & $(5.07\pm0.14)\times10^{-2}$ \\
            & 2007 Dec 26               &                   & XIS3       & 127 ks                 & $(5.17\pm0.08)\times10^{-2}$ \\
            & 2007 Dec 26               &                   & HXD/PIN    & 99 ks                  & $(5.22\pm0.26)\times10^{-2}$ \\ \hline
\Sw         & 2004 Dec 15 - 2006 Nov 01 & SWIFTJ0109.0+1320 & BAT        & 2.8 Ms (effective)     & $(6.31\pm0.89)\times10^{-4}$ \\ \hline
\Ch         & 2005 Nov 08               & 6910              & ACIS-S     & 20 ks                  & $(6.38\pm0.16)\times10^{-2}$ \\ \hline
\Ch         & 2005 Nov 12               & 7200              & ACIS-S     & 20 ks                  & $(6.55\pm0.18)\times10^{-2}$ \\ \hline
\XMM        & 2004 Jan 21               & 0203280301        & EPIC/pn    & 6.3 ks                 & $(1.37\pm0.05)\times10^{-1}$ \\ \hline

\hline
\end{tabular}
\label{obslog}
\end{table}

\begin{table}\small
\centering
\caption{{\sc Fe~K$\alpha$ properties from Model IV}}
\vspace{0.2cm}
\begin{tabular}{ll}
\hline\hline
Parameter & Value \\
\hline
Rest-frame energy & $6.385^{+0.021}_{-0.023}$~keV \\
Line width & $<$65~eV [XIS FWHM = 100eV]\\
Intensity & $(2.40\pm0.04)\times10^{-5}$ photons cm$^{-2}$ s$^{-1}$ \\
Equivalent width &  129$\pm$43~eV \\
\hline
\end{tabular}
\label{fek_parameters}
\end{table}

\clearpage
\newpage

\begin{sidewaystable}
\caption{{\sc Continuum Spectral Parameters for Joint Model Fits to Suzaku and Swift Spectrum. Luminosities quoted are unabsorbed and extrapolated to 2-100 keV.}}
\begin{tabular}{lllllllllll}
\hline\hline
           &      &                                                      &                                                   &                       &                       &                                     &             &             & $L_{\rm (2-100 keV)}$              &                  \\
           &      &                                                      &                                                   &  Absorber             & Covering              &                                     &             & Reflector   & (Power Law)                        &                  \\
Overview & Model  & Description of spectrum                              &  $N_{\rm H}$ (cm$^{-2}$)                          &  log($\xi$)           & fraction              & $\Gamma$                            & $R$         & log($\xi$)  & (ergs s$^{-1}$)                    &  $\chi^{2}$/dof  \\
         & (1)    &              (2)                                     &           (3)                                     &   (4)                 &    (5)                &    (6)                              & (7)        & (8)          &      (9)                           &   (10)           \\ \hline
Single     &  I   & Abs(PL(1)+Gauss)+PL(2)                               &  $(3.7\pm0.20)\times10^{23}$                      &  \nodata              & 100\%                 & $\Gamma_{\rm 1}$=$1.39\pm0.05$      & \nodata     & \nodata     & $(3.9^{+0.5}_{-0.6})\times10^{44}$ & 242/152          \\
absorption &      &                                                      & \nodata                                           &  \nodata              & \nodata               & $\Gamma_{\rm 2}$=$\Gamma_1$         & \nodata     & \nodata     & $(1.7\pm0.1)\times10^{43}$         &                  \\ \hline
Multiple   &  II  & PCAbs(PL+Gauss)                                      & $(3.7\pm0.20)\times10^{23}$                       &  \nodata              & 96$\pm$1\%            & $1.39\pm0.05$                       & \nodata     & \nodata     & $(4.1^{+0.5}_{-0.6})\times10^{44}$ & 242/152          \\
absorption & III  & PCAbs(1)$\times$PCAbs(2)$\times$(PL+Gauss)           & $N_{\rm H,1}$=$(6.2^{+0.9}_{-1.0})\times10^{23}$  &  \nodata              & $84^{+6}_{-10}$ \%    & $1.71\pm0.09$                       & \nodata     & \nodata     & $(4.5^{+1.1}_{-1.6})\times10^{44}$ & 176/150          \\
           &      &                                                      & $N_{\rm H,2}$=$(1.6\pm0.4)\times10^{23}$          &  \nodata              & $91^{+5}_{-3}$ \%     & \nodata                             & \nodata     & \nodata     & \nodata                            &                  \\
           &  IV  & PCAbs(1)$\times$Abs(2)$\times$(PL(1)+Gauss)+PL(2)    & $N_{\rm H,1}$=$(6.0^{+0.9}_{-1.0})\times10^{23}$  &  \nodata              & $85^{+6}_{-4}$ \%     & $\Gamma_{\rm 1}$=$1.71\pm0.09$      & \nodata     & \nodata     & $(4.4^{+1.1}_{-1.6})\times10^{44}$ & 176/150          \\
           &      &                                                      & $N_{\rm H,2}$=$(1.7\pm0.4)\times10^{23}$          &  \nodata              & 100\%                 & \nodata                             & \nodata     & \nodata     & \nodata                            &                  \\
           &      &                                                      & \nodata                                           &  \nodata              & \nodata               & $\Gamma_{\rm 2}$=$\Gamma_1$         & \nodata     & \nodata     & $(6.5\pm0.4)\times10^{42}$         &                  \\ \hline
Warm       &   V  & PCWarmAbs(PL(1)+Gauss)                               & $(5.2^{+0.3}_{-0.2})\times10^{23}$                & $1.51^{+0.17}_{-0.20}$& 98$\pm$1\%            & $\Gamma_{\rm 1}$=1.67$\pm$0.07      & \nodata     & \nodata     & $(4.2^{+0.7}_{-0.1})\times10^{44}$ & 184/151          \\
absorption &      &                                                      & \nodata                                           &  \nodata              & \nodata               & \nodata                             & \nodata     & \nodata     & \nodata                            &                  \\
           &  VI  & WarmAbs(PL(1)+Gauss)+PL(2)                           & $(5.2^{+0.3}_{-0.2})\times10^{23}$                & $1.49^{+0.17}_{-0.20}$& 100\%                 & $\Gamma_{\rm 1}$=1.69$\pm$0.07      & \nodata     & \nodata     & $(4.1^{+0.7}_{-0.8})\times10^{44}$ & 184/151          \\
           &      &                                                      & \nodata                                           &  \nodata              & \nodata               & $\Gamma_{\rm 2}$=$\Gamma_1$         & \nodata     & \nodata     & $(6.9\pm0.4)\times10^{42}$         &                  \\ \hline
Neutral    & VII  & PCAbs(1)$\times$Abs(2)$\times$(RefPL(1)+Gauss)+PL(2) & $N_{\rm H,1}$=$(5.7\pm1.0)\times10^{23}$          &  \nodata              & 86$\pm$6\%            & $\Gamma_{\rm 1}$=$1.75\pm0.08$      & $<$0.41     & \nodata     & $(4.0^{+1.4}_{-1.0})\times10^{44}$ & 175/149          \\
reflection &      &                                                      & $N_{\rm H,2}$=$(1.6^{+0.4}_{-0.6})\times10^{23}$  &  \nodata              & 100\%                 & \nodata                             & \nodata     & \nodata     & \nodata                            &                  \\
           &      &                                                      & \nodata                                           &  \nodata              & \nodata               & $\Gamma_{\rm 2}$=$\Gamma_1$         & \nodata     & \nodata     & $(5.8\pm0.4)\times10^{42}$         &                  \\ \hline
Ionized    & VIII & PCAbs(1)$\times$Abs(2)$\times$(PL(1)+Reflion(2)+Gauss)+PL(3)    & $N_{\rm H,1}$=$(6.6\pm0.2)\times10^{23}$&  \nodata              & 83$\pm$1\%            & $\Gamma_{\rm 1}$=$1.58^{+0.03}_{-0.02}$ & \nodata     & \nodata         & $(1.9\pm0.2)\times10^{44}$ & 174/148         \\ 
reflection &      &                                                                 & $N_{\rm H,2}$=$(2.0\pm0.1)\times10^{23}$&  \nodata              & 100\%                 & $\Gamma_{\rm 2}$=$\Gamma_1$             & \nodata     & 4 (pegged)      & $(2.5\pm0.1)\times10^{44}$ &                 \\
           &       &                                                                 & \nodata                                 &  \nodata              & \nodata               & $\Gamma_{\rm 3}$=$\Gamma_1$             & \nodata     & \nodata         & $(9.6\pm0.5)\times10^{42}$ &                \\ \hline
\label{spectralresults}
\end{tabular}
\begin{minipage}{24cm}
Col. (1): Model number. Col. (2): Description of spectrum (Abs=Neutral absorption, PCAbs=Partially covering neutral absorption, WarmAbs=Warm Absorption ($\xi$ is the ionization parameter and is quoted in units of ergs~cm~s$^{-1}$), PCWarmAbs=Partially covering warm absorption, PL=Power Law, RefPL=Sum of direct and reflected power-law continuum from neutral material, Reflion=\cite{rf05} ionized reflection continuum, Gauss=Redshifted Gaussian line. Col. (3): Intrinsic neutral hydrogen column density. Galactic absorption has also been applied. Col. (4): Ionization parameter of absorber. Col. (5): Covering fraction. Col. (6): Power-law photon index. Col. (7): Reflection fraction. Col. (8): Ionization parameter for \cite{rf05} ionized reflection model. Col. (9): 2--100 keV (extrapolated) unabsorbed luminosity of primary power law. Col. (10): Value of $\chi^{2}$ and degrees of freedom.
\end{minipage}
\end{sidewaystable}

\clearpage

\begin{table}\small
\centering
\caption{Best-fitting parameters for joint fit to Suzaku, Swift, Chandra, and XMM-Newton Spectra. Luminosities quoted are unabsorbed and extrapolated to 2-100 keV.}
\begin{tabular}{lllll}
\hline\hline
                                                 &                                      &                       &                                     & $L_{\rm (2-100 keV)}$                                 \\
                                                 &                                      &  Covering             &                                     & (Power Law)                                           \\
Component                                        &  $N_{\rm H}$ (cm$^{-2}$)             &  fraction             & $\Gamma$                            & (ergs s$^{-1}$)                                       \\ \hline
Partially covering                               & $(9.8^{+3.4}_{-2.3})\times10^{22}$   & 100\%                 & $1.67\pm0.10$                       & $L_{\rm Chandra, 6910}$  = $(3.7^{+1.2}_{-0.8})\times10^{44}$\\
primary power law                                & $(4.9^{+0.6}_{-0.5})\times10^{23}$   & $92^{+3}_{-4}$ \%     & \nodata                             & $L_{\rm Chandra, 7200}$  = $(3.5^{+1.0}_{-0.7})\times10^{44}$\\
                                                 &                                      &                       &                                     & $L_{\rm XMM}$            = $(3.3^{+1.0}_{-0.9})\times10^{44}$\\
                                                 &                                      &                       &                                     & $L_{\rm Suzaku}$         = $(3.8^{+1.2}_{-0.8})\times10^{44}$\\
                                                 &                                      &                       &                                     & $L_{\rm Swift}$          = $(4.1^{+1.9}_{-1.2})\times10^{44}$\\
\hline
Soft power law                                   & \nodata                              & 100\%                 & =$\Gamma_1$                         & $L_{\rm Chandra, 6910}$  = $(2.1\pm0.3)\times10^{42}$\\
                                                 &                                      &                       &                                     & $L_{\rm Chandra, 7200}$  = $(2.5\pm0.4)\times10^{44}$\\
                                                 &                                      &                       &                                     & $L_{\rm XMM}$            = $(3.5\pm0.5)\times10^{44}$\\
                                                 &                                      &                       &                                     & $L_{\rm Suzaku}$         = $(7.0\pm0.5)\times10^{44}$\\ \hline
\label{jointresults}
\end{tabular}
\end{table}

\end{document}